\documentclass{jaa}
\usepackage{natbib}
\usepackage{amsmath}
\usepackage{amssymb}
\usepackage{graphicx}
\usepackage{booktabs}
\usepackage{hyperref}

\begin{document}\sloppy

\title{Compact and Physically Interpretable Feature Models for Photometric Type Ia Supernova Classification}

\author{Anurag Garg\textsuperscript{1, *}}
\affilOne{\textsuperscript{1}Ministry of Education, Abu Dhabi, UAE.\\}

\twocolumn[{

\maketitle

\corres{anurag.garg@moe.sch.ae}

\msinfo{13 March 2026}{}{}

\begin{abstract}

Photometric classification of Type Ia supernovae is essential for modern time-domain surveys, where spectroscopic confirmation is not always feasible for the full transient sample. In this work, we investigate a compact and physically interpretable feature representation derived from multi-band light curves and evaluate its performance using gradient-boosted decision trees on the Supernova Photometric Classification Challenge (SPCC) dataset. The compact representation is derived from an initial pool of 31 light-curve features, reduced to 30 after removing redundant variables, and further optimized to a 16-feature model using systematic ablation and performance analysis. 

The final compact model achieves an F1--score of 0.844 on the held-out test set. This is consistent with k-fold cross-validation results (0.841 $\pm$ 0.006). The precision--recall area under the curve (PR-AUC) is 0.928, with similarly low variance across folds. Relative to our earlier 31-feature optimized XGBoost model for the same SPCC classification task (F1 $\approx$ 0.923), the compact 16-feature representation retains strong classification performance (F1 = 0.844) while substantially improving interpretability and reducing feature-space complexity. The ablation results show that temporal evolution provides the dominant classification signal, while brightness, color, and variability features supply complementary information. A reduced core of approximately ten physically meaningful features retains a large fraction of the performance of the compact model, with only a small decrease in F1-score, indicating that reliable classification does not require large high-dimensional feature spaces.  

These results demonstrate that interpretable feature-based models can capture the essential astrophysical information needed for Type Ia photometric classification, with direct implications for survey cadence, filter coverage, and the design of transparent and efficient machine learning pipelines for future time-domain surveys.

\end{abstract}

\keywords{Supernovae---Data analysis---Statistical---Surveys---Machine Learning.}

}]

\section{Introduction}

Modern time-domain surveys are producing an unprecedented number of transient detections, making automated photometric classification an essential component of supernova research. Type Ia supernovae are of particular importance because of their role as standardizable candles in cosmology \cite{riess1998, perlmutter1999}, and reliable identification of these events is required for constructing large homogeneous samples. However, spectroscopic confirmation is not always possible for all candidates, especially in wide-field surveys where the number of detections exceeds the available follow-up resources. As a result, accurate photometric classification methods have become a central topic in observational astrophysics.

The need for automated classification became evident with the development of large transient surveys and was formalized in the Supernova Photometric Classification Challenge (SPCC), which provided simulated multi-band light curves for testing classification algorithms \cite{kessler2010}. Since then, a wide range of machine learning approaches have been applied to photometric supernova typing, including template fitting, feature-based classifiers, and deep learning models.

Early studies demonstrated that carefully constructed light-curve features can achieve high classification accuracy using relatively simple models \cite{lochner2016}.
Subsequent work explored the use of additional contextual information, such as host-galaxy properties and redshift estimates, to improve performance \cite{villar2019}. More recent approaches have used neural networks and sequence models to classify supernovae directly from light curves without explicit feature extraction \cite{charnock2017, moser2017, burhanudin2022, fraga2024}.
While these methods can achieve strong results, they often require large training sets and produce internal representations that are difficult to relate directly to observable astrophysical quantities.
By contrast, ensemble models trained on explicit light-curve features are generally easier to interpret because the input variables have clear physical meaning, and the contribution of individual features can be examined through feature importance rankings, split behaviour, or post-hoc explanation tools \cite{garg2025optimizing}.

It is important to contextualize the present results relative to prior work on the same dataset. Optimized feature-based ensemble models on SPCC have reported higher absolute performance (e.g., F1 $\gtrsim$ 0.90 and PR-AUC $\gtrsim$ 0.97) when using larger feature sets and extensive hyperparameter tuning \cite{lochner2016, burhanudin2022, fraga2024, garg2025optimizing}. In particular, our prior optimized XGBoost model achieved F1 $\approx$ 0.923 on the same dataset \cite{garg2025optimizing}. The objective of the present work is therefore not to maximize absolute performance but to explicitly quantify the trade-off between performance and interpretability under strong feature compression.  

Interpretability is particularly important for survey-scale applications, where classifiers trained on one dataset may be applied to another survey with different cadence, noise, or filter systems. Models that rely on physically meaningful observables may generalize better across surveys than models based on high-dimensional latent representations.

Photometric measurements from different surveys can differ in several instrumental and observational aspects, even when the same filters are nominally used. These differences include variations in filter transmission curves, detector sensitivity, photometric calibration procedures, cadence and sampling strategy, signal-to-noise characteristics, and treatment of background and host-galaxy light. Such factors can lead to systematic shifts in measured fluxes, color indices, and temporal sampling of the light curve. As a result, features derived from photometric data may not be directly comparable across surveys unless they are constructed from physically meaningful and robust observables. Understanding these instrumental differences is therefore important when assessing the generalization of classification models trained on one survey to another \cite{villar2019}.

An alternative approach is therefore to construct compact feature sets derived directly from observable properties of the light curve, such as peak brightness, color indices, variability statistics, and temporal evolution. These quantities have clear astrophysical meaning and can be measured from multi-band photometry without detailed model fitting. Using physically motivated features may improve interpretability, reduce overfitting, and make the classifier more robust to changes in survey configuration.

Although compact feature representations have been used in several studies, the relative importance of different types of observables has not been systematically quantified. In particular, it is not clear which physical aspects of the light curve carry the strongest discriminating power for distinguishing Type Ia supernovae from other transient classes, and how classification performance changes when specific features are removed.

Previous studies have demonstrated that feature-based models can achieve high classification accuracy in photometric supernova classification when physically motivated light-curve descriptors are used as classifier inputs \cite{karpenka2013, lochner2016, burhanudin2022, fraga2024, garg2025optimizing}. However, the physical contribution of individual light-curve descriptors has rarely been examined in a controlled and systematic way. In particular, explicit feature-ablation studies based on physically interpretable observables derived directly from multi-band light curves remain relatively uncommon in the supernova-classification literature. As a result, it remains unclear which measurable properties of the light curve are essential for reliable photometric classification, which are redundant, and which aspects of the classification signal depend on survey cadence, filter coverage, or temporal sampling. Addressing this gap is important both for understanding the astrophysical origin of the classification performance and for designing robust pipelines for future large-scale time-domain surveys.

In this work, we investigate photometric Type Ia supernova classification using a compact and physically interpretable feature representation derived from multi-band light curves. The compact 16-feature representation used in this work is derived from a larger feature set introduced in \cite{garg2025optimizing}, which contained 31 candidate features constructed from multi-band light-curve statistics. A reduced working set of 30 features was first obtained after removing redundant or unstable variables. From this pool, a compact subset of 16 features was selected based on systematic ablation experiments and performance comparison. The reduction was guided by empirical evaluation rather than heuristic selection, and preserves a large fraction of the classification performance while significantly reducing dimensionality. The classifier is based on gradient-boosted decision trees, which provide strong performance together with feature-level interpretability.

To understand which observables are most important, we perform a systematic feature ablation analysis. Individual features and physically related groups of features are removed, and the resulting change in performance is measured using F1-score and precision--recall AUC (PR-AUC). We also construct progressively smaller feature subsets to identify the minimal set of variables required for reliable classification.

This approach makes it possible to interpret the classifier in physical terms. Rather than treating the model as a black box, the ablation results reveal how brightness, color, variability, and temporal evolution contribute to the decision boundary. This allows the behaviour of the classifier to be connected directly to the known properties of thermonuclear supernova explosions.

Understanding the physical origin of the classification signal also has practical implications for survey design and survey operation. The importance of cadence, filter coverage, and temporal sampling can be evaluated by measuring how the removal of corresponding features affects performance. Such analysis may help identify which observing conditions are most important for reliable photometric typing in ongoing and forthcoming time-domain surveys such as DES, ZTF, and LSST \cite{lsst2009, des2005, ztf2019}. In practice, the observing strategy of a major survey is usually optimized for multiple science goals and cannot be redefined solely for supernova classification. However, results of this kind can still inform follow-up prioritization, rolling-cadence choices where available, filter-allocation trade-offs, and the design of downstream classification pipelines that must operate under the survey's existing sampling pattern.

The paper is organized as follows. Section~2 describes the dataset used in this work. Section~3 presents the classification model and evaluation metrics, including the training protocol and model-selection procedure. Section~4 introduces the compact feature representation. Section~5 summarizes the baseline performance of the compact model. Section~6 presents the feature ablation experiments and their physical interpretation. Section~7 discusses the implications for survey design and feature engineering. Section~8 outlines the main limitations of the present study and possible directions for future work. Section~9 summarizes the conclusions.

\section{Dataset}

The analysis in this work is performed using the Supernova Photometric Classification Challenge (SPCC) dataset \cite{kessler2010}. The SPCC was designed as a community benchmark for photometric supernova classification and consists of simulated DES-like multi-band light curves in the $g$, $r$, $i$, and $z$ filters with realistic cadence, observing conditions, and measurement noise. The challenge was constructed to mimic the practical situation in which only a subset of discovered supernovae can be spectroscopically confirmed, while the majority must be classified photometrically. This makes the SPCC a relevant test bed for methods intended for large time-domain surveys.

The processed binary dataset used in this work contains 21,319 objects in total, of which 5,088 are labeled as Type~Ia and 16,231 as non-Ia. In the original multiclass simulation, the non-Ia sample includes several core-collapse subclasses, namely II, IIL, IIP, IIn, Ib, Ibc, and Ic. The compact-feature workflow therefore treats the task as binary Ia versus non-Ia classification rather than full transient taxonomy. This is an important limitation: the present methodology is designed to distinguish thermonuclear supernovae from other supernova classes represented in the SPCC and does not explicitly address non-supernova variable sources such as AGN, variable stars, or tidal-disruption events, which would need to be incorporated through broader training data in a survey-deployment setting.

Each object in the SPCC is represented by an irregularly sampled light curve with observation times, flux measurements, and flux uncertainties in the four photometric bands. The number of observations per object varies across the dataset, reflecting realistic survey conditions. In the processed compact-feature table used here, the observation count ranges from 16 to 161 per object, with a median of 101, and the observed time span ranges from 30 to 174.907 days, with a median of 125.852 days. The simulated redshift spans 0.023 to 1.1, with a median of 0.6655. These statistics summarize the effective sample used for feature extraction.

For the experiments reported in this paper, the data are divided using a fixed stratified split. The training set contains 13,644 objects (3,256 Ia and 10,388 non-Ia), the validation set contains 3,411 objects (814 Ia and 2,597 non-Ia), and the held-out test set contains 4,264 objects (1,018 Ia and 3,246 non-Ia). The class fractions are therefore preserved to high accuracy across all splits. The combined training-plus-validation set contains 17,055 objects, of which 4,070 are Ia and 12,985 are non-Ia.

From the raw light curves, a set of summary features is extracted for each object. These features describe the brightness scale, color evolution, flux variability, and temporal behaviour of the transient. The feature extraction procedure is designed to use only observable quantities that can be computed directly from the photometric data, without fitting a physical template model. Because the present workflow operates on flux-based engineered features, the processed compact table does not directly store a raw peak-magnitude summary; accordingly, the manuscript reports dataset size, class balance, redshift range, and observation statistics as the primary dataset descriptors.

\begin{figure}[!t]
\centering
\includegraphics[width=0.95\columnwidth]{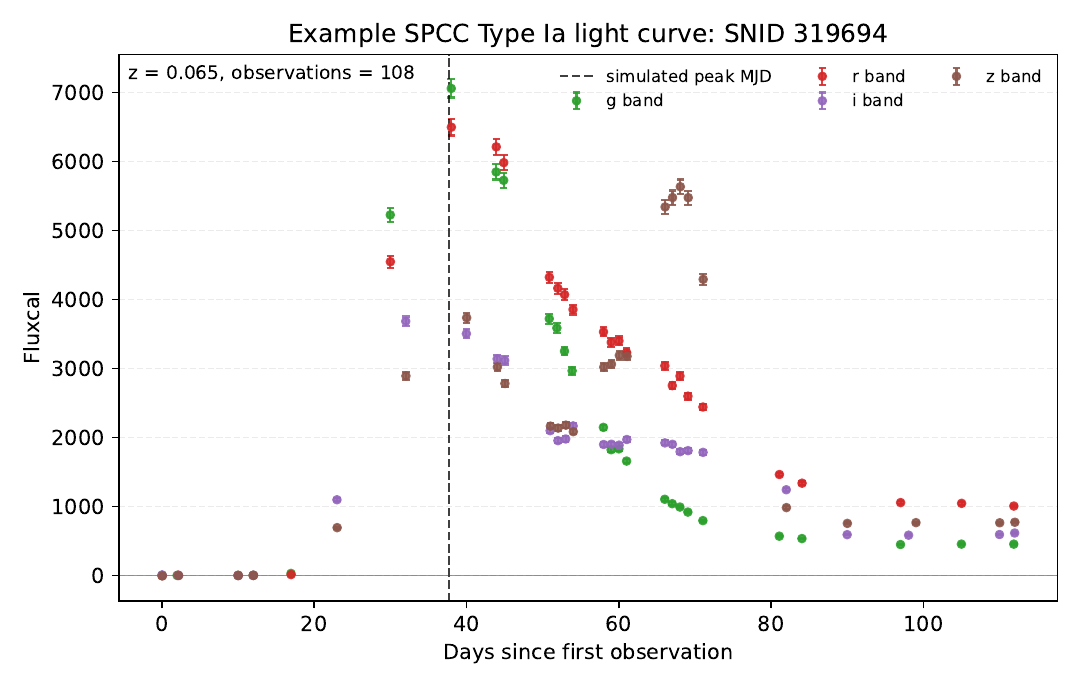}
\caption{Example SPCC Type Ia multi-band light curve for SNID 319694 ($z = 0.065$; 108 observations) in the $g$, $r$, $i$, and $z$ bands. Points show flux measurements with associated uncertainties. The dashed vertical line indicates the simulated peak epoch.}
\label{fig:example_lc}
\end{figure}

To illustrate the nature of the data, a representative multi-band light curve from the SPCC sample is shown in Figure~\ref{fig:example_lc}. The example highlights the irregular cadence, measurement uncertainties, and band-dependent temporal evolution that characterize photometric supernova observations.

\section{Model and Evaluation Metrics}

The classification model is implemented using gradient-boosted decision trees, which provide strong performance for tabular feature data while allowing the importance of individual variables to be evaluated. Gradient boosting is well suited for photometric classification because it can capture non-linear relationships between features without requiring a large number of training samples.

The model is trained using the extracted light-curve features under a fixed train/validation/test protocol. The data are first split with a held-out test fraction of 0.2, and 0.2 of the remaining train+validation pool is then used for validation, giving an effective split of 64\% training, 16\% validation, and 20\% test. The train and validation partitions are used for model selection, after which the final classifier is refit on the combined training and validation data and evaluated once on the held-out test partition. This protocol allows a direct comparison between the compact model and the reduced feature subsets studied in the ablation analysis.

Class imbalance is handled through \texttt{scale\_pos\_weight}, computed as the ratio of non-Ia to Ia objects in the current training partition.

The data split used in all experiments is stratified with 64\% training, 16\% validation, and 20\% held-out test data, ensuring consistent class fractions across partitions. Hyperparameters are selected from the grid shown in Table~\ref{tab:xgb_grid}. The compact model uses a learning rate $\eta \in [0.03, 0.05]$, tree depth $\in [3, 5]$, subsample and colsample\_bytree fractions in the range $0.8$--$0.9$, minimum child weight $1$--$2$, and $\ell_2$ regularization $\lambda$ in the range $1.0$--$1.5$. The number of boosting rounds is determined using early stopping (patience of 30 rounds based on validation logloss), typically converging within $\sim$150--250 trees.

Before fitting, the input features are standardized using feature-wise mean and standard deviation computed from the training partition, and the same transformation is applied to validation and test data. The baseline classifier used throughout the compact-feature workflow is XGBoost with objective \texttt{binary:logistic}, evaluation metric \texttt{logloss}, tree method \texttt{hist}, maximum 400 boosting rounds during model selection, and early stopping after 30 rounds without validation improvement.

\begin{table}[!t]
\centering
\caption{XGBoost hyperparameter grid used for compact-model selection.}
\label{tab:xgb_grid}
\setlength{\tabcolsep}{3pt}
\begin{tabular}{ccccccc}

\hline

ID & depth & $\eta$ & subsample & \shortstack{colsample\\\_bytree} & \shortstack{min\_child\\\_weight} & $\lambda$ \\

\hline

1 & 3 & 0.05 & 0.8 & 0.8 & 1.0 & 1.0 \\

2 & 4 & 0.05 & 0.9 & 0.9 & 1.0 & 1.0 \\

3 & 5 & 0.03 & 0.8 & 0.8 & 2.0 & 1.5 \\

\hline

\end{tabular}
\vspace{2pt}

{\footnotesize
\parbox{0.9\columnwidth}{
Note. depth = \texttt{max\_depth}; $\eta$ = \texttt{learning\_rate}; sub. = \texttt{subsample}; col. = \texttt{colsample\_bytree}; child = \texttt{min\_child\_weight}; $\lambda$ = \texttt{reg\_lambda} (L2 regularization strength).
}
}
\end{table}

Model performance is evaluated using two primary metrics:

\begin{itemize}

\item F1-score, which provides a balance between precision and recall and is appropriate when the classes are not equally represented.

\item Precision--recall AUC (PR-AUC), which is particularly useful for imbalanced datasets because it directly measures the trade-off between sample purity and completeness without being dominated by the large number of true negatives \cite{Hastie2009}.

\end{itemize}

Validation metrics are used for model selection, with PR-AUC taken as the primary criterion because of the class imbalance in the Ia versus non-Ia task. The validation set is used only during hyperparameter selection and early stopping and is not used for final performance reporting.

After model selection, the final classifier is retrained on the combined training and validation data, and all reported performance metrics are computed on the held-out test set. This ensures that the reported results reflect unbiased generalization performance rather than model-selection behaviour. For the compact 16-feature baseline, the test-set values are Accuracy = 0.916745, Precision = 0.762887, Recall = 0.944990, F1 = 0.844, ROC-AUC = 0.976588, and PR-AUC = 0.927761. During boosting, validation logloss is monitored for early stopping; thus, the training workflow tracks both predictive accuracy summaries and the corresponding validation-loss evolution, even though only the final metric values are emphasized in the present manuscript. Using the same model and evaluation procedure for each experiment ensures that the effect of feature removal can be interpreted directly in terms of the physical information contained in the light-curve descriptors.

\begin{figure}[!t]
\centering
\includegraphics[width=0.95\columnwidth]{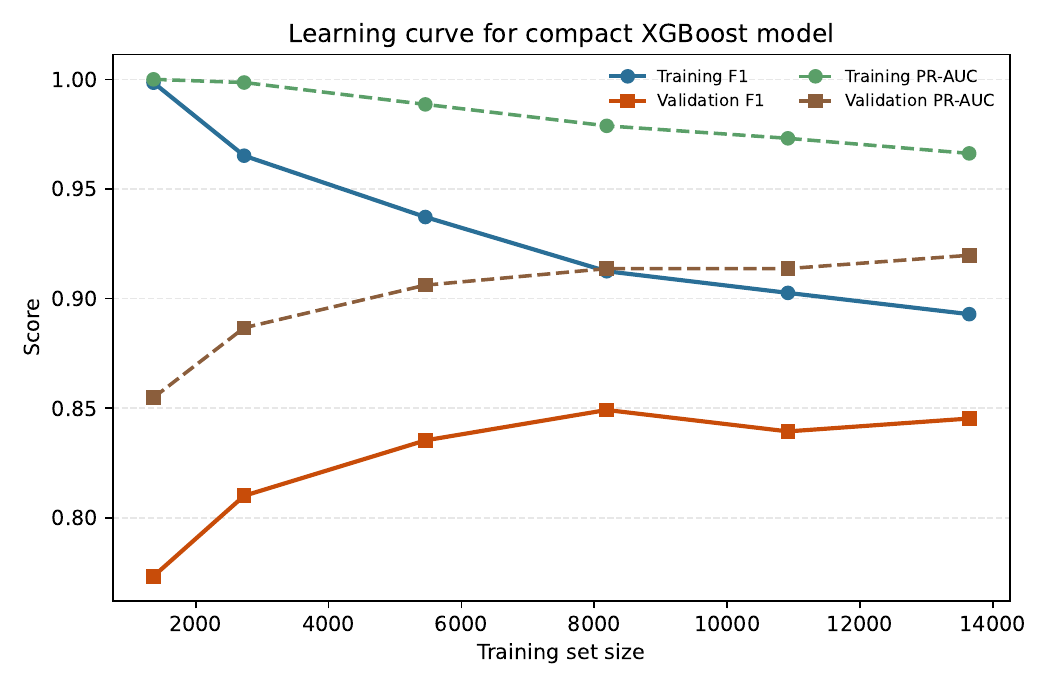}
\caption{
Learning curve for the compact XGBoost model showing training and validation performance as a function of training set size. The convergence of training and validation F1-score indicates that the model generalizes well and does not exhibit strong overfitting.
}
\label{fig:learning_curve}
\end{figure}

To further assess model behaviour during training, both learning curves and convergence diagnostics were evaluated. Figure~\ref{fig:learning_curve} shows the variation of training and validation F1-score as a function of training set size. At small sample sizes, the model exhibits a clear gap between training and validation scores, indicating a modest gap between training and validation performance. As the training set increases, the validation F1-score improves and approaches the training performance, demonstrating that the model benefits from additional data and that the compact feature representation captures stable predictive structure.

\begin{figure}[!t]
\centering
\includegraphics[width=0.95\columnwidth]{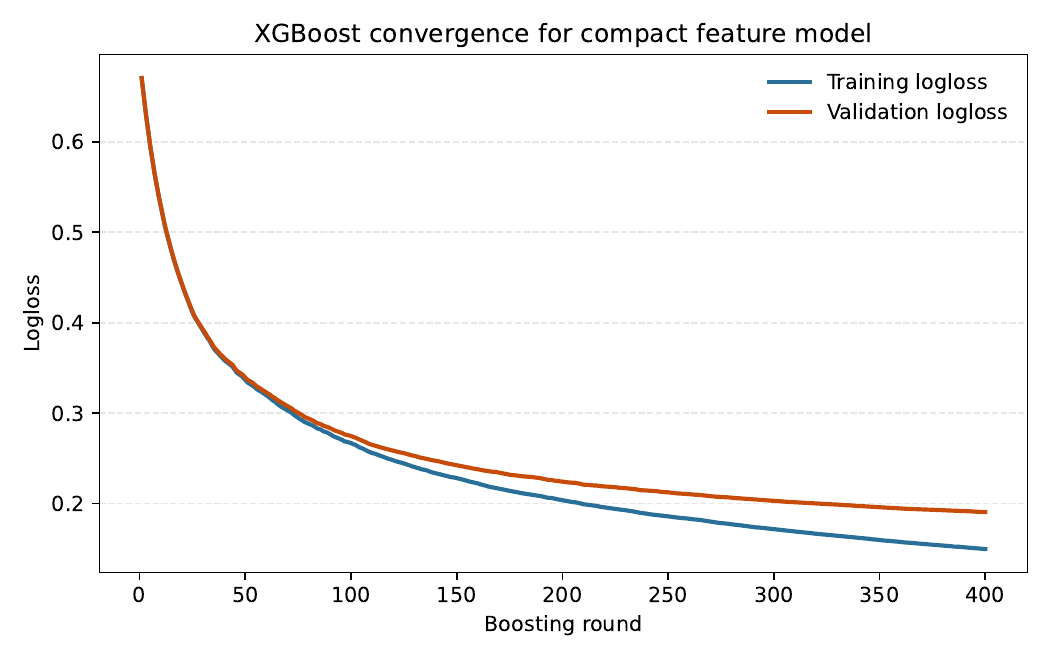}
\caption{
Training and validation logloss as a function of boosting rounds for the compact model. The smooth decrease and eventual stabilization of validation loss indicate stable optimization and effective convergence.
}
\label{fig:logloss_curve}
\end{figure}

Convergence behaviour during boosting is illustrated in Figure~\ref{fig:logloss_curve}, which shows the evolution of training and validation logloss across boosting rounds. Both curves decrease smoothly with no evidence of divergence, and the validation loss stabilizes toward later iterations. This indicates well-behaved optimization and supports the use of early stopping as an effective regularization mechanism.

These results confirm that the compact model is not only accurate at the final evaluation stage, but also exhibits stable training dynamics and consistent generalization behaviour.

\section{Compact Feature Model}

Photometric classification methods often rely on large feature sets or high-dimensional representations of the light curve. While such approaches can achieve high performance, they may reduce interpretability and increase the risk of overfitting to a particular dataset. For survey-scale applications, it is desirable to construct a compact feature representation that preserves the essential physical information while keeping the model simple and interpretable.

In this work, we use a feature set derived directly from observable properties of the multi-band light curve. The features are computed from the flux measurements in the $g$, $r$, $i$, and $z$ bands and describe brightness, color, variability, and temporal evolution. These quantities include peak flux, mean flux, flux dispersion, peak time in each band, color proxies constructed from flux ratios, and the total observational time span of the light curve. The compact feature set is summarized in Table~\ref{tab:feature_list}. A detailed description of the feature construction, including mathematical definitions, transformations, and implementation details, is provided in Appendix~A and in the associated repository \cite{GargRepo2026}.

The compact feature representation consists of 16 explicitly defined variables spanning brightness, color, variability, and temporal evolution. These features were selected from a larger pool of 31 candidate features introduced in \cite{garg2025optimizing}. A working set of 30 features was first constructed, after which systematic feature-ablation experiments were used to identify a minimal subset that preserves classification performance. Table~\ref{tab:feature_list} lists the final 16 features used in the model, and Appendix~A provides their precise mathematical definitions.

The compact-feature pipeline consists of the following feature-construction steps:

\begin{itemize}
\item \textbf{Observation preparation:} Load the raw multi-band photometry, sort observations by MJD and band, and normalize times relative to the first observation.
\item \textbf{Basic cleaning:} remove measurements with non-finite MJD, flux, or flux uncertainty values, and reject events with invalid metadata or too few valid observations.
\item \textbf{Shared-grid construction:} form a common event-level time grid from all unique observation times without interpolation, Gaussian-process reconstruction, or template fitting.
\item \textbf{Feature summarization:} compute band-wise brightness, color, variability, and temporal summary statistics from the reconstructed observed grid.
\end{itemize}

The feature-extraction pipeline operates directly on the observed SPCC photometry and uses a deliberately simple reconstruction procedure. For each event, the observations are first sorted by Modified Julian Date (MJD) and band, and individual measurements are removed only when MJD, flux, or flux uncertainty is non-finite. Events are rejected only if required metadata are missing or invalid, or if too few valid observations remain after this filtering. No explicit sigma-clipping, host-contamination correction, or upper-limit treatment is applied in the compact-feature pipeline.

To construct band-comparable summaries, all unique observation times for an event are collected into a shared time grid defined by the normalized coordinate
\begin{equation}
 t_j = \mathrm{MJD}_j - \mathrm{MJD}_{\mathrm{first}},
\end{equation}
where no additional time rescaling is applied. For each band $b \in \{g,r,i,z\}$, the observed flux is used at times when that band is measured, and missing band values on the shared grid are filled with zero flux and zero flux uncertainty. The compact-feature pipeline therefore does not use Gaussian processes, spline interpolation, or template fitting; peak times and other summary statistics are determined directly on this reconstructed observed grid.

For each band $b$, the peak index is defined as
\begin{equation}
 j_{\mathrm{peak},b} = \arg\max_j F_b(t_j),
\end{equation}
with peak flux and peak time given by
\begin{equation}
 F_{\mathrm{peak},b} = \max_j F_b(t_j), \qquad t_{\mathrm{peak},b} = t_{j_{\mathrm{peak},b}}.
\end{equation}
Peak-flux features are stored after log compression as $\log_{10}(1 + \max(F_{\mathrm{peak},b},0))$, while peak-time features are stored directly as the normalized time coordinate $t_{\mathrm{peak},b}$. Mean-flux features use signed-log compression, standard-deviation and amplitude features use log-compressed positive summaries, and the total time span is defined as $\max_j t_j - \min_j t_j$. Color features are magnitude-style peak-flux-ratio proxies,
\begin{equation}
 C_{a-b} = -2.5\log_{10}\left(\frac{F_{\mathrm{peak},a}}{F_{\mathrm{peak},b}}\right),
\end{equation}
computed using a positive flux floor of $10^{-6}$ and clipped to the range $[-5,5]$ to avoid unstable extreme ratios. These color variables should therefore be interpreted as bounded photometric proxies rather than direct spectroscopic colors.

Flux uncertainties are loaded from the raw data, but they are not propagated through the compact-feature extraction by Monte Carlo sampling or analytic error propagation. Similarly, the present pipeline does not explicitly model upper limits, strong host-galaxy contamination, or outlier rejection beyond the removal of non-finite measurements. These simplifications improve transparency and reproducibility, but they also define an important limitation of the present compact-feature study.

\begin{table}[htb]
\centering
\caption{Compact 16-feature set used in this work. This table provides a high-level summary of the physically interpretable features; precise mathematical definitions are given in Appendix~A.}
\label{tab:feature_list}
\small
\setlength{\tabcolsep}{3pt}
\begin{tabular}{@{}p{0.44\columnwidth}p{0.44\columnwidth}@{}}
\toprule
Feature & Physical Interpretation \\
\midrule
$z\_\text{peak\_flux}$ & z-band peak flux ($z_\text{peak\_flux}$) \\
$r\_\text{mean\_flux}$ & Average brightness scale in $r$ band \\
$g\_\text{mean\_flux}$ & Average brightness scale in $g$ band \\
$r\_\text{peak\_flux}$ & Maximum brightness in $r$ band \\
$i\_\text{peak\_flux}$ & Maximum brightness in $i$ band \\
\midrule
$C_{g-r}$ & Spectral slope between $g$ and $r$ \\
$C_{r-i}$ & Spectral slope between $r$ and $i$ \\
$C_{i-z}$ & Red-end spectral behaviour \\
\midrule
$i\_\text{std\_flux}$ & Variability in $i$ band \\
$z\_\text{std\_flux}$ & z-band flux variability ($z_\text{std\_flux}$) \\
$r\_\text{std\_flux}$ & Variability in $r$ band \\
$i\_\text{amplitude}$ & Light-curve amplitude in $i$ band \\
\midrule
$r\_\text{time\_of\_peak}$ & Peak epoch in $r$ band \\
$i\_\text{time\_of\_peak}$ & Peak epoch in $i$ band \\
$z\_\text{time\_of\_peak}$ & Peak epoch in $z$ band \\
$\text{time\_span}$ & Duration of observed light curve (not rise/decline time) \\
\bottomrule
\end{tabular}
\end{table}

The classifier is implemented using gradient-boosted decision trees, which provide strong performance while allowing feature-level interpretation. The model is trained on the SPCC dataset using the compact feature set, and performance is evaluated using F1-score and precision--recall AUC.

The compact model achieves high classification accuracy while using only a small number of physically motivated variables. This indicates that reliable photometric typing does not require large feature spaces, but can be obtained from a limited set of observable light-curve properties.

To examine which features contribute most strongly to the decision, we compute SHAP (SHapley Additive exPlanations) values for the trained model. SHAP values provide a measure of the contribution of each feature to the classification output for individual objects, allowing the importance of different physical descriptors to be visualized.

\begin{figure}[!t]
    \centering
        \includegraphics[width=0.95\columnwidth]{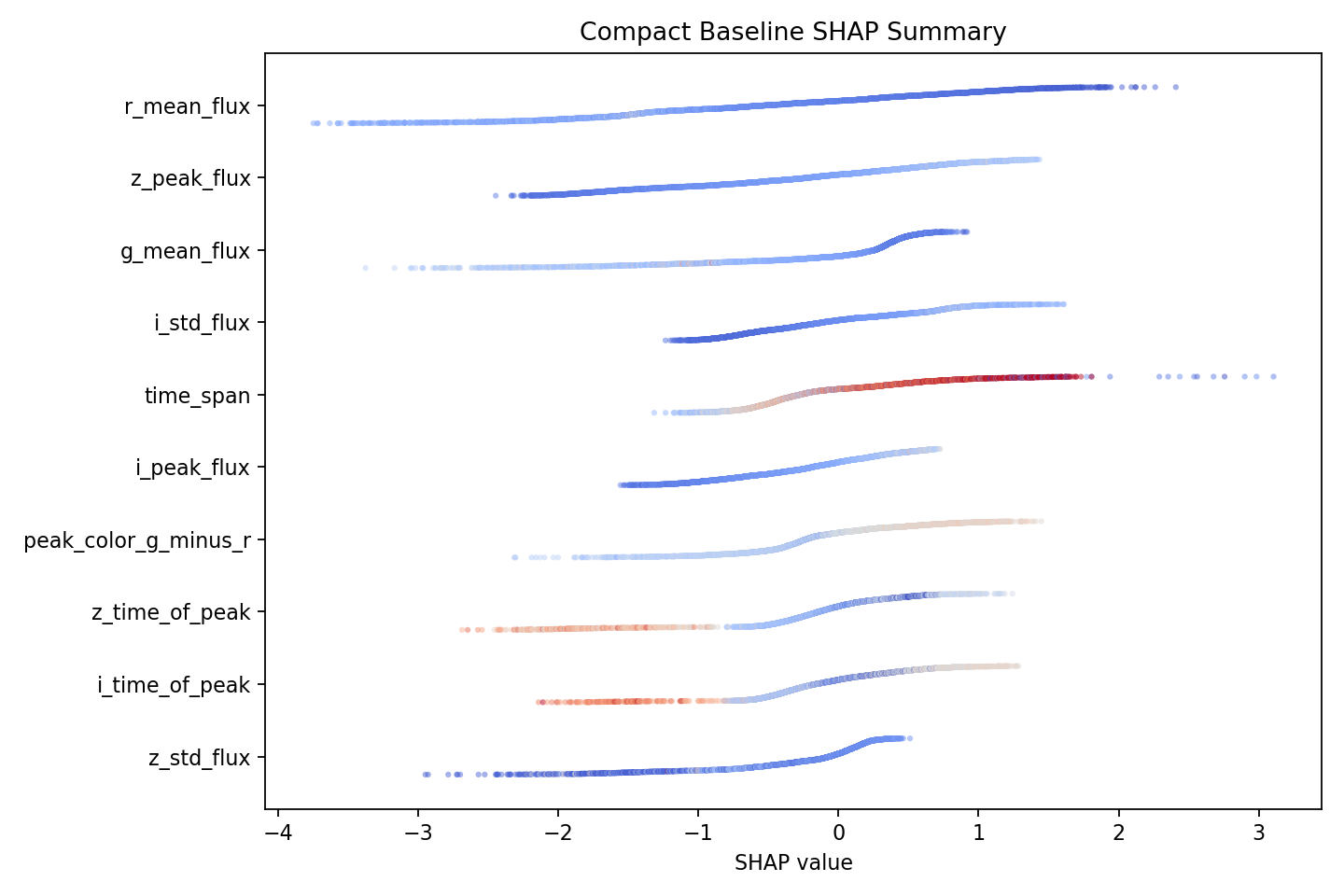}
        \caption{
        SHAP summary plot for the compact feature model. Only the top-ranked features are shown for clarity, although the full model contains 16 features. Each point represents an individual object, with horizontal position indicating the contribution to the model output and color representing the feature value (red = high, blue = low). Temporal features and z-band flux statistics provide the largest contribution to the classification decision, indicating that the multi-band time evolution of the light curve is the dominant source of discriminating information.
        }
        \label{fig:shap}
\end{figure}

The resulting SHAP summary plot is shown in Figure~\ref{fig:shap}. The plot displays the top-ranked features based on their average contribution to the model output. Although the compact model consists of 16 features, only the most important features are shown for clarity, following standard SHAP visualization practice. The reduced set of approximately ten features shown in the figure corresponds closely to the minimal core identified in the ablation analysis, but is selected purely for visualization and does not represent a separately trained model.

The remaining features have smaller average SHAP contributions and are omitted for visual clarity; their inclusion does not alter the qualitative ranking.

In the SHAP summary plot, each point represents an individual object, and the horizontal position indicates the contribution of that feature to the model prediction. The color scale represents the feature value, with red indicating higher values and blue indicating lower values. This allows the relationship between feature magnitude and classification impact to be visualized directly.

As seen in Figure~\ref{fig:shap}, temporal descriptors such as the total time span and peak times have the strongest influence on the model, followed by brightness and color features. This suggests that the classifier relies primarily on the multi-band temporal structure of the light curve, which is consistent with the known behaviour of Type Ia supernovae.

\section{Model Performance}

The performance of the compact feature model is evaluated primarily using precision--recall behaviour and the resulting F1-score. These metrics provide a robust assessment of classification quality, particularly in the presence of class imbalance, which is common in photometric supernova datasets.

For the compact 16-feature model, the test-set performance reaches an F1-score of 0.844 and a PR-AUC of 0.928. These values provide the baseline reference for all subsequent ablation experiments.

\begin{figure}[!t]
\centering
\includegraphics[width=0.95\columnwidth]{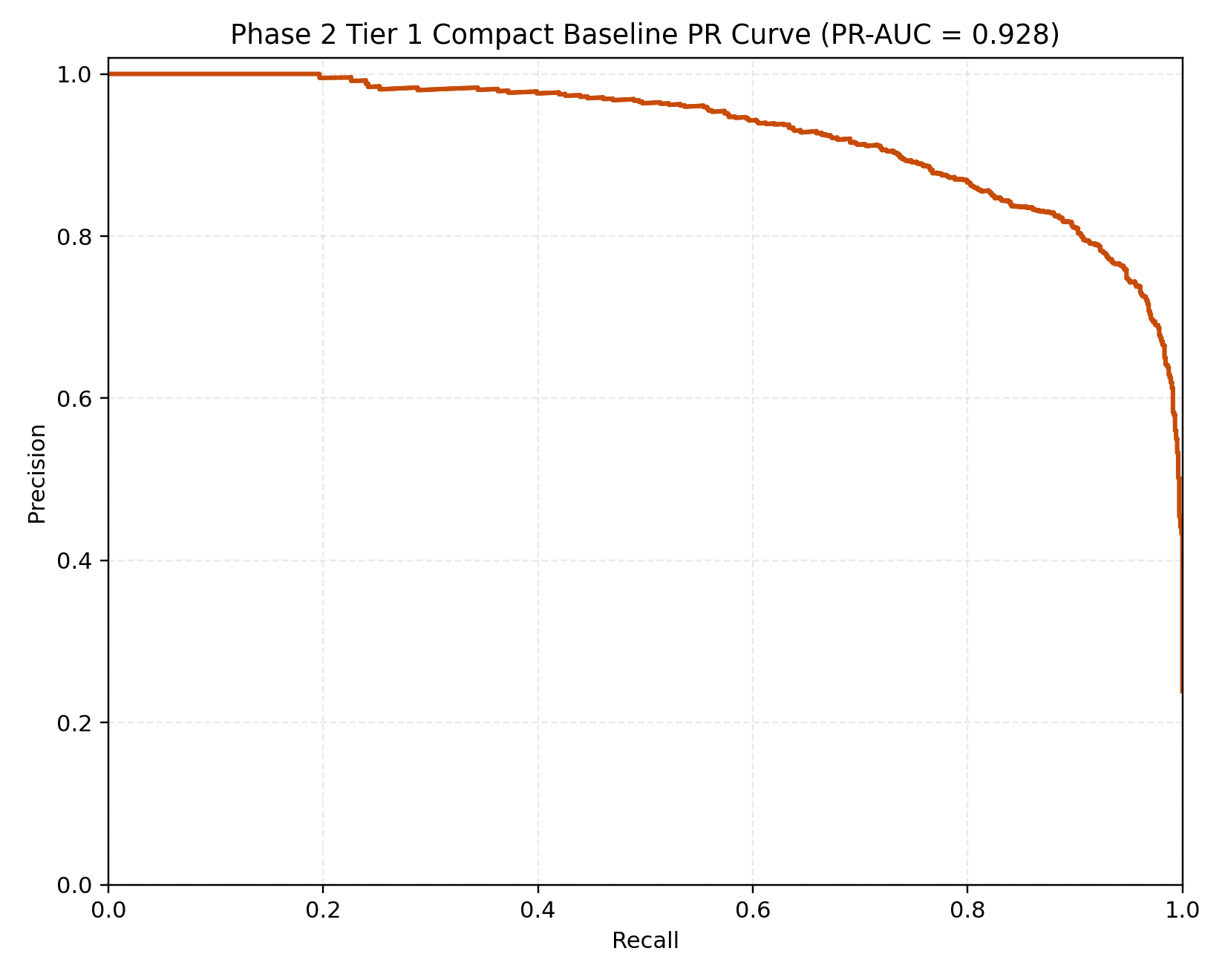}
\caption{
Precision--recall curve for the compact feature model.
The model maintains high precision over a wide range of recall,
indicating reliable identification of Type Ia supernovae.
}
\label{fig:pr}
\end{figure}

The precision--recall curve measures the trade-off between completeness and purity of the Type Ia sample. High precision indicates that objects classified as Type Ia are reliable, while high recall ensures that most true Type Ia events are recovered. The precision--recall curve for the compact model is shown in Figure~\ref{fig:pr}.

As seen in Figure~\ref{fig:pr}, the classifier preserves high precision over a wide range of recall values, indicating that most objects identified as Type Ia are genuine while retaining a large fraction of the true Type Ia population.
The numerical values reported above make clear that the compact feature representation achieves strong classification performance even without resorting to high-dimensional feature spaces or deep neural architectures.

This focus on F1-score and PR-AUC is motivated by the class imbalance of the SPCC dataset, for which ROC-based summaries can be less informative because they are strongly influenced by the large number of true negatives. The high scores obtained with the compact feature set demonstrate that accurate photometric classification can be achieved without using large feature spaces or deep neural networks. This motivates a more detailed investigation of which physical features are essential for the classification, which is explored in the next section using feature ablation.

\begin{table}[!t]
\centering
\caption{Comparison of compact feature model with larger feature sets.}
\label{tab:feature_comparison}
\small
\setlength{\tabcolsep}{5pt}
\begin{tabular}{lcccc}
\toprule
Feature Set & Count & F1 & ROC-AUC & PR-AUC \\
\midrule
Full baseline & 31 & 0.837 & 0.976 & 0.928 \\
Working set & 30 & 0.840 & 0.976 & 0.928 \\
Compact baseline & 16 & 0.844 & 0.976 & 0.928 \\
\bottomrule
\end{tabular}
\end{table}

As shown in Table~\ref{tab:feature_comparison}, the compact 16-feature model preserves a large fraction of the discriminative performance of the larger feature sets. While the PR-AUC shows only a negligible decrease relative to the full baseline, the F1-score improves, indicating a better balance between precision and recall at the operating threshold. This confirms that the reduction in dimensionality does not degrade classification quality and supports the use of the compact representation as an efficient baseline.

The full baseline (31 features) corresponds to the feature set introduced in \cite{garg2025optimizing}.

\begin{table*}[!t]
\centering
\caption{Contextual comparison with representative photometric supernova classification results from the literature. Reported ranges correspond to results obtained on SPCC or comparable simulated survey datasets. Literature ranges are summarized from representative feature-based and deep-learning studies cited in the final column.}
\label{tab:sota_spcc}
\small
\setlength{\tabcolsep}{5pt}
\begin{tabular}{p{0.29\textwidth}p{0.20\textwidth}p{0.10\textwidth}p{0.10\textwidth}p{0.23\textwidth}}
\toprule
Study class & Feature type & F1 & PR-AUC & Reference \\
\midrule
Feature-based (representative literature) & engineered & $\sim$0.90--0.93 & $\sim$0.97--0.99 & \cite{lochner2016, burhanudin2022} \\
Deep learning (representative literature) & sequence/image & $\sim$0.90+ & $\sim$0.98+ & \cite{charnock2017, moser2017, fraga2024} \\
Prior optimized work & engineered (31+) & $\sim$0.92 & $\sim$0.97+ & \cite{garg2025optimizing} \\
This work (compact) & 16 features ($\sim$10 core) & 0.844 & 0.928 & this study \\
\bottomrule
\end{tabular}
\end{table*}

Table~\ref{tab:sota_spcc} summarizes the positioning of the present work relative to representative cited studies on SPCC. While absolute performance is lower than highly optimized or high-dimensional models, the compact representation retains a large fraction of the discriminative power with substantially improved interpretability and reduced feature complexity.

Although direct comparisons are affected by differences in feature construction, training protocols, and evaluation splits across studies, the results summarized in Table~\ref{tab:sota_spcc} provide a representative context for the performance–interpretability trade-off explored in this work.

\begin{table}[!t]
\centering
\caption{Cross-validation stability of the compact model under different resampling protocols. Mean values are computed across resampled runs and are distinct from the held-out test-set performance reported elsewhere.}
\label{tab:cv_stability}
\small
\setlength{\tabcolsep}{4pt}
\begin{tabular}{lccccc}
\toprule
Protocol & Runs & Mean F1 & SD & ROC-AUC & PR-AUC \\
\midrule
k-fold CV & 5 & 0.841 & 0.006 & 0.975 & 0.923 \\
Rand. split & 5 & 0.845 & 0.003 & 0.976 & 0.923 \\
\bottomrule
\end{tabular}
\end{table}

As shown in Table~\ref{tab:cv_stability}, the compact model exhibits consistent performance across different resampling protocols, with low variance in F1-score, indicating that the results are not sensitive to a particular train/test split.

\section{Feature Ablation and Physical Interpretation of the Compact Model}

To understand which physical properties of the light curve drive the performance of the compact 16-feature model described in the previous section, we performed a systematic feature ablation study. The goal of this analysis is not only to measure performance degradation when features are removed, but also to identify which physical aspects of the photometric evolution carry the most discriminative power for Type Ia supernova classification.

The ablation experiments were performed at three levels:

\begin{enumerate}
\item single-feature removal,
\item block-level removal of physically related feature groups, and
\item progressive subset construction to determine the minimal feature core.
\end{enumerate}

All experiments were evaluated using F1-score and PR-AUC, with the compact feature model used as the baseline reference.

\subsection{Single-Feature Ablation}

\begin{table}[htb]
    \tabularfont
    \caption[]{Single-feature ablation results for the compact model. Negative values indicate degradation in F1-score after removal of the corresponding feature.}
    \label{tab:single_ablation}
    \begin{tabular}{p{0.30\columnwidth}cp{0.4\columnwidth}}
        \topline
        Feature & $\Delta$F1 & Interpretation \\
        \hline
        Time span & -0.034 & total temporal extent of light curve \\
        $z$-band flux dispersion & -0.021 & variability in $z$ band \\
        $z$-band peak flux & -0.019 & z-band peak brightness \\
        $i$-band time of peak & -0.017 & peak timing in $i$ band \\
        Peak color $g-r$ & -0.012 & color index $g-r$ \\
        Peak color $r-i$ & -0.010 & color index $r-i$ \\
        Peak color $i-z$ & -0.009 & color index $i-z$ \\
        $r$-band peak flux & +0.002 & redundant brightness scale \\
        $i$-band peak flux & +0.001 & redundant brightness scale \\
        \hline
    \end{tabular}
\end{table}

Each feature was removed individually while keeping all others fixed. Table~\ref{tab:single_ablation} lists the change in F1-score after removing each feature from the compact model. The largest performance degradation is observed when removing the global temporal descriptor, the total time span of the light curve, indicating that the overall duration of the observed light curve is the single most informative variable in the compact representation.

Moderate performance drops are observed when removing the $z$-band flux dispersion, the $z$-band peak flux, and the time of peak in the $i$ band. These features describe variability in the $z$ band, peak brightness in the $z$ band, and the timing of the peak in the $i$ band. Their importance suggests that both multi-band timing and z-band flux behaviour contain strong class-discriminating information.

Color features such as the peak color $g-r$, the peak color $r-i$, and the peak color $i-z$ produce smaller but consistent performance reductions. Although color encodes important physical information related to temperature and its evolution, its comparatively smaller impact in this model suggests that color alone is not sufficient to uniquely distinguish Type Ia supernovae from other classes. This is likely because different supernova types can exhibit overlapping color evolution at certain phases, especially when observational noise and redshift effects are present. Similar trends have been reported in feature-based studies, where temporal and multi-band information contribute strongly to classification performance, although these effects are not always isolated explicitly from color-based features \cite{lochner2016, burhanudin2022}.

A small number of features, including the $r$-band peak flux and the $i$-band peak flux, have negligible or slightly positive effects when removed, implying partial redundancy within the compact feature set.

\subsection{Block-Level Ablation}

\begin{figure}[!t]
\centering
\includegraphics[width=0.95\columnwidth]{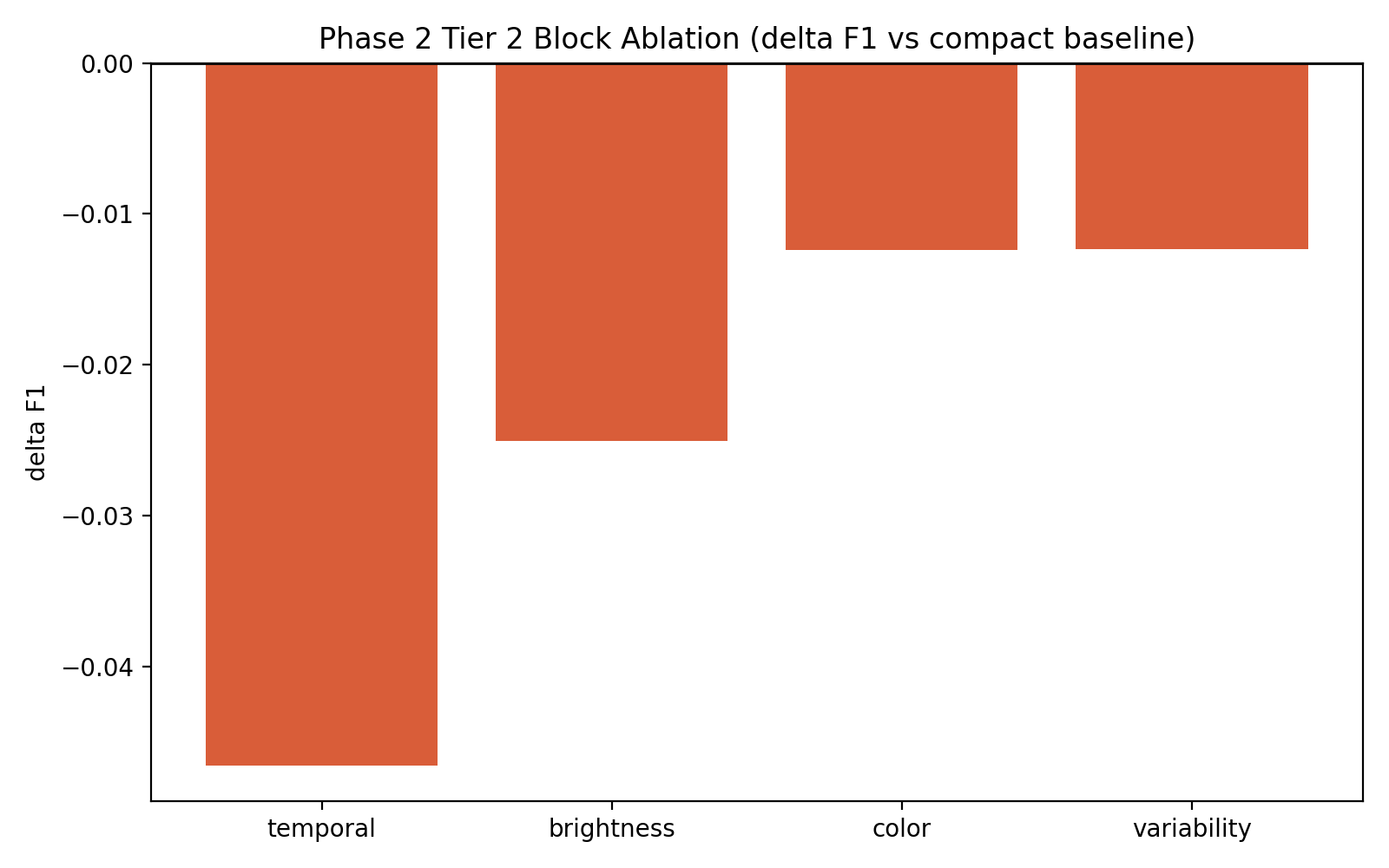}
\caption{
Change in F1-score when groups of physically related features
are removed.
Temporal features cause the largest performance drop,
followed by brightness, color, and variability.
This indicates that the time evolution of the light curve
is the dominant physical signature used by the classifier.
}
\label{fig:ablation_block}
\end{figure}

To better understand the physical origin of the classification power, the features are grouped into blocks corresponding to different observable properties of the light curve. These blocks represent temporal evolution, brightness scale, color information, and flux variability. Each block is removed in turn, and the model is retrained to measure the change in performance. The result of this block-level ablation test is shown in Figure~\ref{fig:ablation_block}.

\begin{itemize}
\item Brightness features (peak and mean flux values)
\item Color features (band differences)
\item Variability features (flux dispersion statistics)
\item Temporal features (time span and peak epochs)
\end{itemize}

As seen in Figure~\ref{fig:ablation_block}, removing the temporal feature block produces the largest degradation in performance, confirming that the time structure of the light curve is the most important discriminator between Type Ia and other supernova types. This is expected because Type Ia events exhibit a characteristic temporal evolution across photometric bands, including systematic differences in peak timing and overall light-curve duration compared to core-collapse supernovae. Although explicit rise and decline rates are not directly used as features in the present model, these aspects are indirectly reflected in temporal descriptors such as the total time span and band-dependent peak times.

Brightness features also have a strong effect, reflecting the relatively narrow luminosity distribution of Type Ia supernovae. Color features produce a smaller but noticeable decrease, indicating that spectral evolution contributes to the classification. Variability-related features have the smallest impact, suggesting that they provide secondary information once the temporal and brightness properties are known.

The resulting hierarchy of importance is therefore

\begin{equation}
\mathrm{Temporal} > \mathrm{Brightness} > \mathrm{Color} \approx \mathrm{Variability}.
\end{equation}

This ordering is physically reasonable, since Type Ia supernovae are characterized not only by their luminosity but by a highly structured temporal evolution across filters.

\subsection{Subset Growth and Information Hierarchy}

To examine how classification performance builds as physical information is added, models were trained using progressively larger subsets of features.

\begin{figure}[!t]
\centering
\includegraphics[width=0.95\columnwidth]{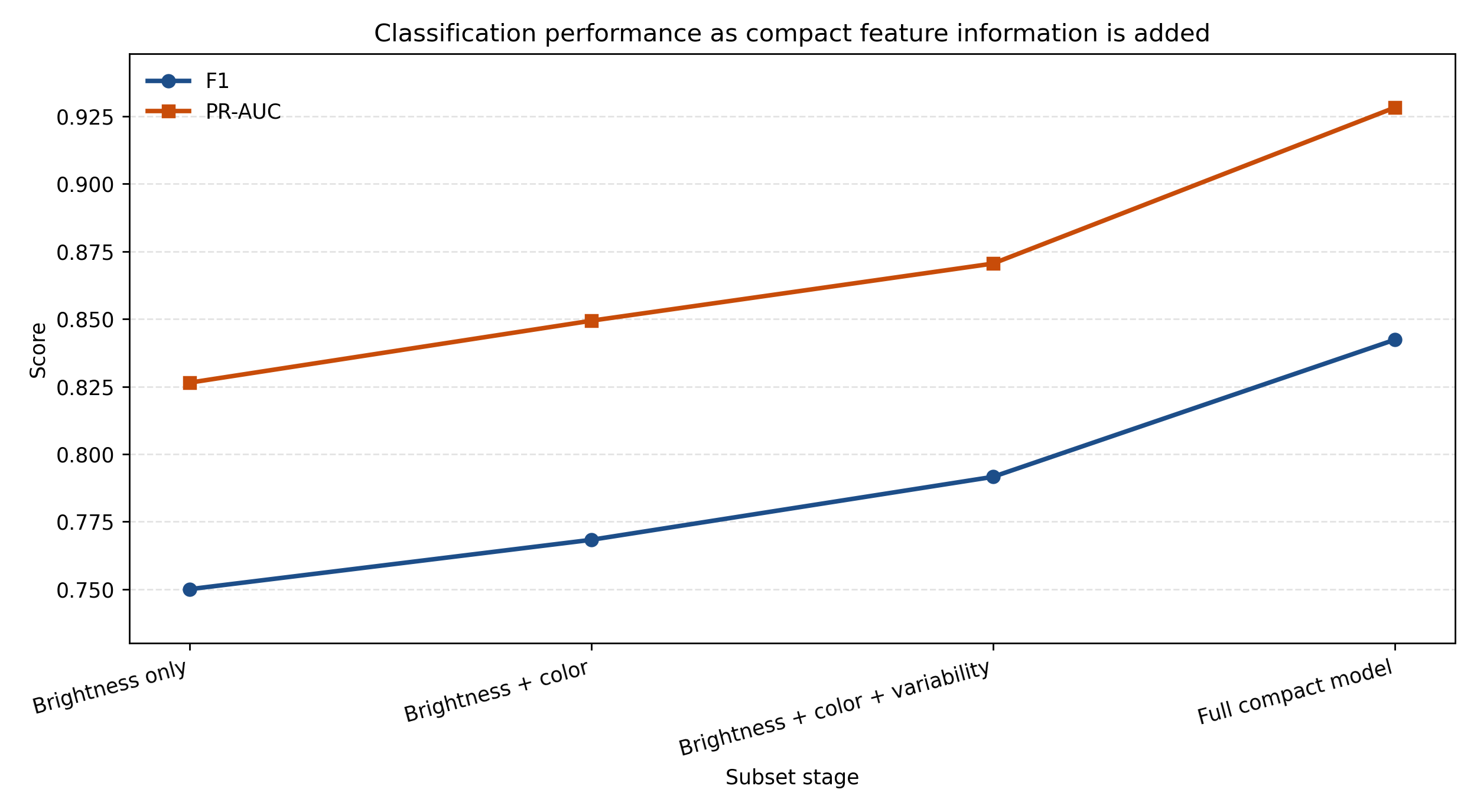}
\caption{
Classification performance as progressively larger subsets of physically motivated features are included. Brightness alone provides a first-order
separation between classes, while the largest improvement occurs when temporal descriptors are added, indicating that multi-band time evolution
provides the dominant discriminating information.
}
\label{fig:subset_growth}
\end{figure}

As shown in Figure~\ref{fig:subset_growth}, using only brightness features yields moderate performance (F1 $\approx 0.75$), indicating that luminosity alone carries significant signal. Adding color features produces only a small improvement, while adding variability descriptors gives a modest additional gain.

The largest improvement occurs when temporal features are included, raising the F1-score to the full compact-model value. This demonstrates that temporal information acts as the decisive disambiguation signal once brightness and color provide an initial separation.

This behaviour indicates that the classification problem is interaction-driven, with the strongest discrimination arising from the joint structure of brightness, color, and time evolution.

\subsection{Minimal Core Feature Set}

\begin{figure}[!t]
\centering
\includegraphics[width=0.95\columnwidth]{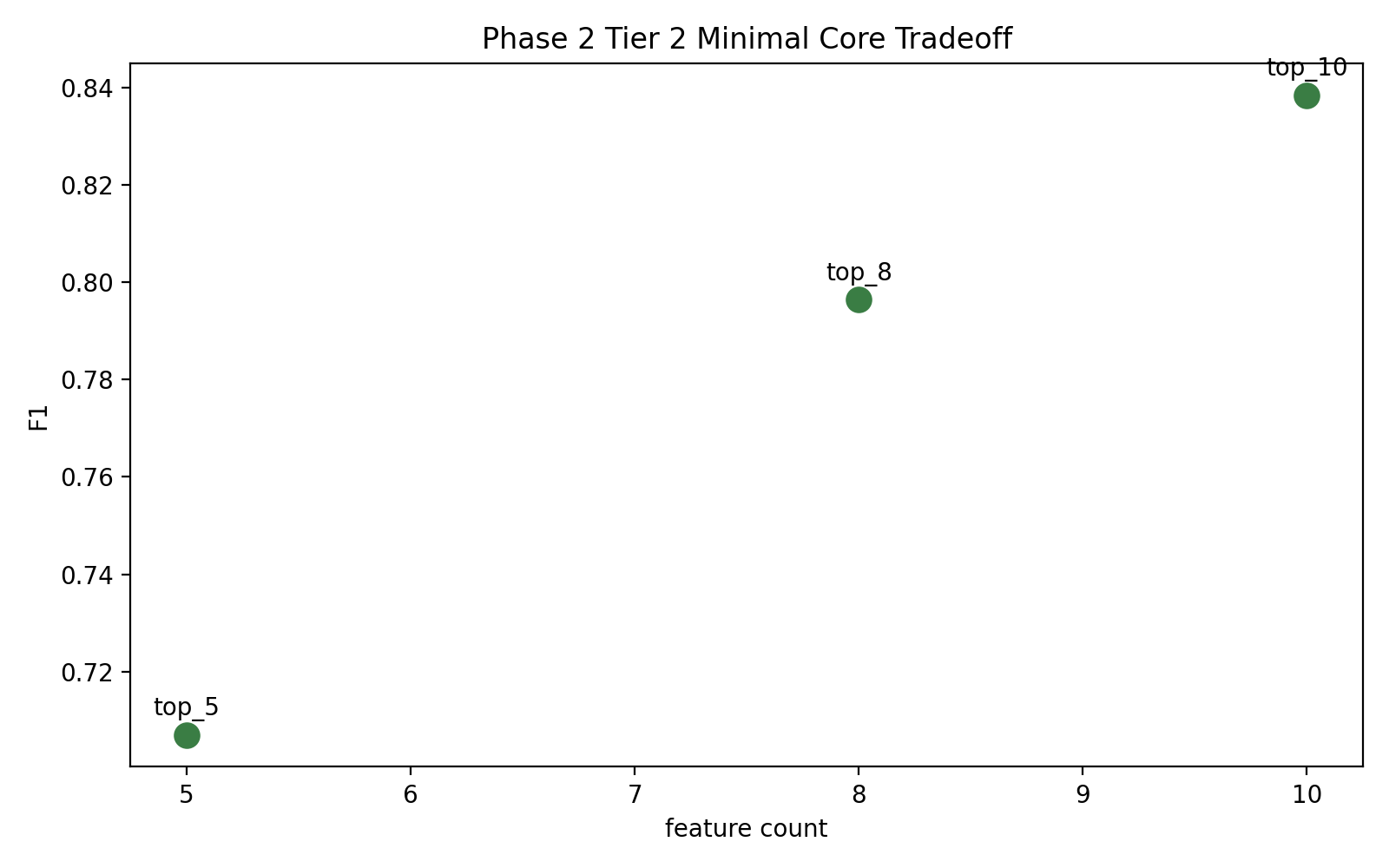}
\caption{
Trade-off between the number of features and classification performance. A reduced feature set containing only the most important variables retains a large fraction of the performance of the compact model, indicating that photometric classification depends on a small number of physically meaningful descriptors.
}
\label{fig:minimal_core}
\end{figure}

To determine whether the compact model contains redundant information, we constructed reduced feature sets using the most important features identified from the ablation ranking. Models were trained with progressively smaller subsets to identify the minimal number of variables required to retain the performance of the compact feature model. The relation between feature count and classification performance is shown in Figure~\ref{fig:minimal_core}.

As shown in Figure~\ref{fig:minimal_core}, a subset containing approximately ten features retains a large fraction of the performance of the 16-feature compact model, with a decrease in F1-score of only 0.0057, while smaller subsets show progressively larger degradation. The retained variables include one z-band brightness scale, color indices, a variability statistic, the global time span, and peak times in multiple filters.

\begin{table}[!t]
\centering
\caption{Performance of reduced feature subsets relative to the compact model.}
\label{tab:minimal_core_table}
\begin{tabular}{lcccc}
\toprule
Subset & Count & F1 & PR-AUC & $\Delta$F1 vs 16 \\
\midrule
top 5 & 5 & 0.706973 & 0.745903 & -0.137257 \\
top 8 & 8 & 0.796399 & 0.874361 & -0.047830 \\
top 10 & 10 & 0.838509 & 0.916900 & -0.005721 \\
\bottomrule
\end{tabular}
\end{table}

Table~\ref{tab:minimal_core_table} quantifies the trade-off between feature count and classification performance. The top-10 subset retains nearly the full performance of the 16-feature model, while smaller subsets show progressively larger degradation. This demonstrates that the classification signal is concentrated in a compact set of physically meaningful descriptors and that a compact core can capture most of the relevant information.

These quantities correspond closely to the fundamental physical descriptors of a photometric transient:

\begin{itemize}
\item luminosity scale,
\item spectral shape,
\item temporal duration,
\item band-dependent peak evolution,
\item flux variability.
\end{itemize}

The existence of such a compact core indicates that reliable Type Ia classification does not require large feature sets but instead depends on a limited number of physically interpretable observables that describe the brightness, color, and time evolution of the explosion.

\subsection{Astrophysical Interpretation}

The ablation results provide a physically interpretable picture of photometric supernova classification.

Brightness determines the overall scale of the event, color describes the spectral energy distribution, and variability reflects light-curve shape. However, the strongest discriminating power comes from the multi-band temporal evolution.

This is consistent with the known physics of Type Ia supernovae, which follow a relatively uniform thermonuclear explosion sequence, producing characteristic rise times, peak ordering across filters, and decline behaviour that differs from most core-collapse events.

Therefore, the compact model appears to rely primarily on physically meaningful descriptors of the explosion evolution, as indicated by the feature-level interpretability and ablation behaviour.

These results demonstrate that a small, interpretable feature set can capture the essential astrophysical information needed for reliable Type Ia photometric classification.

\section{Implications for Survey Design and Feature Engineering}

The feature ablation and subset growth experiments provide insight not only into the behaviour of the classifier, but also into the type of observational information required for reliable photometric supernova classification. Because the compact model is built from physically motivated light-curve descriptors, the importance ranking of features can be interpreted directly in terms of survey sampling, filter coverage, and temporal cadence.

\subsection{Dominance of Temporal Information}

The block-level ablation shows that temporal features produce the largest performance degradation when removed. In particular, the global time span and band-dependent peak times are among the most critical variables in the compact model.

This result implies that reliable classification depends strongly on how well the temporal evolution of the transient is sampled. A survey that measures only a few isolated flux points may recover the peak brightness but will not capture the temporal structure of the light curve, including the overall duration and the relative timing of flux evolution across different bands. Although explicit rise and decline rates are not directly used as features in the present model, these aspects are indirectly encoded through quantities such as the total time span and band-dependent peak times. These temporal descriptors are found to be essential for distinguishing Type Ia supernovae from other transient classes.

From an observational standpoint, this suggests that cadence is at least as important as depth for classification purposes.

A moderately deep survey with good temporal coverage may outperform a deeper survey with sparse sampling when the goal is photometric typing under fixed observational resources. However, in practice, survey design must balance both detection and classification requirements. Deep observations are essential for detecting high-redshift Type Ia supernovae, which are critical for cosmological studies. The present result should therefore be interpreted as highlighting the importance of cadence for classification performance, rather than suggesting that survey depth is unimportant. An optimal survey strategy must jointly consider depth, cadence, and multi-band coverage to enable both reliable detection and accurate photometric classification.

\subsection{Role of Multi-Band Observations}

Color features and z-band flux statistics show consistent, although smaller, contributions to the classification performance. This confirms that multi-band photometry provides important spectral information that cannot be replaced by single-band measurements.

The ablation results indicate that the reddest bands ($i$ and $z$ in the present feature set) are particularly informative. These bands are sensitive to the cooling phase of the ejecta and to the shift of the spectral energy distribution toward longer wavelengths as the supernova evolves.

This behaviour is expected from the physics of Type Ia explosions, where the temperature decreases after maximum light, producing predictable changes in the relative flux between filters. Capturing this evolution requires observations in multiple bands, preferably extending into the red or near-infrared.

\subsection{Brightness as First-Order Separation}

Brightness features remain important but are not dominant. Removing peak flux and mean flux values causes a noticeable, but smaller, performance loss compared to removing temporal information.

This suggests that luminosity provides a first-order separation between classes, but is insufficient for reliable classification by itself. Different supernova types can overlap in brightness, especially when redshift effects and observational noise are present.

Therefore, classification models that rely primarily on magnitude-based cuts are expected to be less robust than those that incorporate temporal and color information.

\subsection{Existence of a Compact Physical Core}

One of the most significant results of the subset growth experiment is the existence of a small feature subset that preserves nearly the full performance of the compact model.

The minimal core contains approximately ten features describing

\begin{itemize}
\item overall duration of the event,
\item peak timing in multiple bands,
\item one z-band brightness scale,
\item color differences,
\item one variability statistic.
\end{itemize}

These quantities correspond closely to the fundamental physical observables of a thermonuclear explosion observed through broadband photometry.

The fact that such a small set is sufficient suggests that photometric Type Ia classification does not require large, high-dimensional feature spaces, but instead depends on a limited number of physically meaningful measurements.

This has practical consequences for survey pipelines, where reducing the number of required features can simplify data processing and improve interpretability.

\subsection{Consequences for Future Time-Domain Surveys}

Large time-domain surveys such as DES, ZTF, and LSST are already producing vast numbers of transient alerts, and will continue to expand these data volumes further, making spectroscopic confirmation of all candidates impossible. In this context, photometric classification must be both accurate and computationally efficient.

The present results suggest that reliable classification can be achieved with a compact set of interpretable features, provided that the survey delivers

\begin{itemize}
\item sufficient temporal sampling,
\item multi-band coverage,
\item measurements extending beyond peak brightness.
\end{itemize}

These requirements are consistent with the design goals of modern wide-field surveys, but the ablation results highlight that loss of cadence or loss of filter coverage may have a direct impact on classification accuracy.

Therefore, feature-level analysis of classification models can be used as a diagnostic tool for evaluating survey strategies, in addition to its role in improving machine learning performance.

\subsection{Interpretability and Physical Consistency}

An important advantage of the compact feature approach is that each input variable corresponds to a physically interpretable quantity. The ablation study suggests that the model relies primarily on features that have clear astrophysical meaning, such as peak time, color index, and light-curve duration.

This may reduce the risk that the classifier is exploiting dataset-specific correlations that would not generalize to other surveys. Instead, the decision boundary appears to be based on properties that reflect the underlying explosion physics.

As a result, the compact model is expected to be more robust when applied to different data sets, filter systems, or observing conditions.
This expectation is based on the physical interpretability of the features and is not directly validated here; explicit cross-survey testing is required to confirm this behaviour.

This property is essential for future survey-scale applications, where the training and deployment data may not be identical.

\section{Limitations and Future Work}

Although the compact feature model and the ablation experiments provide a clear picture of which physical descriptors are most important for photometric Type Ia supernova classification, several limitations must be considered when interpreting the present results.

\subsection{Dependence on the SPCC Simulation}

All experiments in this work are based on the Supernova Photometric Classification Challenge (SPCC) dataset. While this dataset is widely used and provides realistic light-curve sampling, it remains a simulated survey with fixed assumptions about noise, cadence, and filter response.

A classifier optimized on a single simulation may learn correlations that are specific to that survey configuration. For example, the relative importance of red-band features or temporal coverage may depend on the cadence and depth assumed in the simulation.

Future work should repeat the same ablation analysis on data sets generated with different survey parameters, or on real survey data in order to verify that the feature hierarchy remains stable. Direct cross-survey validation remains an important next step and is planned in future work.

\subsection{Sensitivity to Sampling and Missing Data}

The present feature set assumes that peak times, flux statistics, and color indices can be measured reliably. In practice, real survey light curves often contain missing observations, irregular cadence, and incomplete filter coverage.

Temporal features were found to be the most important variables in the compact model, which implies that classification performance may degrade significantly when the light curve is sparsely sampled. This effect may be particularly important for high-redshift events, where the signal-to-noise ratio is lower and the observed time span is shorter.

Future work should investigate the robustness of the compact feature set under incomplete or noisy observations, and determine whether alternative features can compensate for missing data.

A systematic investigation of cross-survey generalization, including application to real survey data, is beyond the scope of the present study. No claim of transferability is made here; such validation is deferred to future work.

\subsection{Lack of Host-Galaxy and Redshift Information}

The current model uses only light-curve features and does not include host-galaxy properties or redshift information. In real survey pipelines, host-galaxy context is often available and can provide additional constraints on the transient type.

Previous studies have shown that including host-galaxy features can improve classification accuracy, particularly when photometric information alone is ambiguous.

However, host-galaxy features may introduce survey-dependent biases, since galaxy catalogs differ between surveys in several ways. These differences include survey depth, which affects host-galaxy detectability and completeness; image resolution, which influences the measurement of morphological properties such as size, shape, and structural classification; photometric calibration and filter systems, which impact derived quantities such as galaxy colors and stellar mass estimates; and source-detection pipelines, which determine how host associations are assigned. As a result, host-galaxy features derived from one survey may not be directly comparable to those from another, potentially introducing systematic biases when models are transferred across surveys. The compact feature model intentionally avoids such information to preserve generality.

Future work should evaluate whether host-galaxy and redshift features can be incorporated without compromising the interpretability and portability of the model.

\subsection{Model Dependence and Algorithm Choice}

This work focuses on gradient-boosted decision trees, which provide strong performance while allowing feature-level interpretation. Different machine learning algorithms may rely on different combinations of features, even when trained on the same data.

Deep neural networks, for example, can learn complex representations directly from the light curves, but often require large training sets and provide less transparent decision rules.

The compact feature hierarchy identified here should therefore be regarded as model-dependent to some extent. Future studies should test whether the same feature ranking appears when using alternative classifiers, including neural networks, random forests, and probabilistic models.

\subsection{Generalization Across Surveys}

A key requirement for photometric classification is the ability to generalize across different instruments and survey strategies. The present work evaluates performance using a single data distribution, which does not fully test cross-survey robustness.

In practice, a classifier trained on one survey may encounter different noise levels, filter systems, or cadence patterns when applied to another survey. If the model relies on survey-specific correlations, classification accuracy may degrade.

Future work should train and test the compact feature model on mixtures of simulated surveys, or on data from multiple real surveys in order to determine whether the physically motivated feature set remains stable under domain shift.

\subsection{Future Directions}

The results of the ablation analysis suggest several directions for further investigation:

\begin{itemize}
\item evaluation of the compact feature set on multiple survey simulations,
\item robustness tests with incomplete or irregular light curves,
\item inclusion of host-galaxy and redshift information,
\item comparison with deep learning approaches,
\item application to real survey data.
\end{itemize}

These extensions will help determine whether the compact, physically interpretable feature set identified in this work can serve as a general framework for photometric supernova classification in future large-scale time-domain surveys.

\section{Conclusion}

We presented a compact, physically interpretable feature representation for photometric Type Ia supernova classification based on multi-band light-curve descriptors. Using a gradient-boosted decision-tree model, the 16-feature representation achieves strong performance on the SPCC dataset (F1 $\approx 0.844$, PR-AUC $\approx 0.928$).

Feature-ablation experiments show that temporal descriptors (overall duration and band-dependent peak times) provide the dominant discriminating signal, with brightness offering first-order separation and color/variability contributing secondary information. A reduced core of $\sim$10 features retains nearly the full performance, indicating that reliable classification can be achieved with a small set of physically meaningful observables.

These results suggest that effective photometric typing depends primarily on cadence, multi-band coverage, and sampling beyond peak. While demonstrated on a single simulated survey, the approach provides a transparent and interpretable baseline for robust, survey-scale classification pipelines. Future work will test stability across surveys, robustness to incomplete sampling, and extensions with contextual information.

\appendix

\section{Compact Feature Definitions}

An explicit mathematical description of each of the 16 features is provided in Table~\ref{tab:feature_appendix}. In the notation below, $F_b(t_j)$ denotes the reconstructed grid flux in band $b$ at shared time $t_j$, $F_{\mathrm{peak},b}$ denotes the band peak flux, $\bar{F}_b$ denotes the mean band flux, $\sigma_b$ denotes the bandwise flux standard deviation, and $t_0$ denotes the first observed epoch of the event. The associated repository \cite{GargRepo2026} provides the exact implementation used in this work, including transformation details, clipping rules, and the full feature-construction pipeline.

\begin{table*}[htb]
\centering
\caption{Detailed definition of the compact feature set.}
\label{tab:feature_appendix}
\small
\setlength{\tabcolsep}{4pt}
\begin{tabular}{p{0.16\textwidth}p{0.36\textwidth}p{0.36\textwidth}}
\toprule
Feature & Definition & Interpretation \\
\midrule
$z\_\text{peak\_flux}$ & $\log_{10}(1 + F^{\max}_{z})$ & z-band peak flux ($z_\text{peak\_flux}$) \\
$r\_\text{mean\_flux}$ & $\mathrm{sign}(\bar{F}_r)\,\log_{10}(1 + |\bar{F}_r|)$ & Average brightness scale in $r$ band \\
$g\_\text{mean\_flux}$ & $\mathrm{sign}(\bar{F}_g)\,\log_{10}(1 + |\bar{F}_g|)$ & Average brightness scale in $g$ band \\
$r\_\text{peak\_flux}$ & $\log_{10}(1 + F^{\max}_{r})$ & Maximum brightness in $r$ band \\
$i\_\text{peak\_flux}$ & $\log_{10}(1 + F^{\max}_{i})$ & Maximum brightness in $i$ band \\
\midrule
$C_{g-r}$ & $\mathrm{clip}\!\left[-2.5\log_{10}\!\left(\frac{\max(F_{\mathrm{peak},g},10^{-6})}{\max(F_{\mathrm{peak},r},10^{-6})}\right),-5,5\right]$ & Bounded color proxy between $g$ and $r$ \\
$C_{r-i}$ & $\mathrm{clip}\!\left[-2.5\log_{10}\!\left(\frac{\max(F_{\mathrm{peak},r},10^{-6})}{\max(F_{\mathrm{peak},i},10^{-6})}\right),-5,5\right]$ & Bounded color proxy between $r$ and $i$ \\
$C_{i-z}$ & $\mathrm{clip}\!\left[-2.5\log_{10}\!\left(\frac{\max(F_{\mathrm{peak},i},10^{-6})}{\max(F_{\mathrm{peak},z},10^{-6})}\right),-5,5\right]$ & Bounded red-end color proxy \\
\midrule
$i\_\text{std\_flux}$ & $\log_{10}(1 + \sigma_i)$ & Variability in $i$ band \\
$z\_\text{std\_flux}$ & $\log_{10}(1 + \sigma_z)$ & z-band flux variability ($z_\text{std\_flux}$) \\
$r\_\text{std\_flux}$ & $\log_{10}(1 + \sigma_r)$ & Variability in $r$ band \\
$i\_\text{amplitude}$ & $\log_{10}(1 + (F^{\max}_i - F^{\min}_i))$ & Light-curve amplitude in $i$ band \\
\midrule
$r\_\text{time\_of\_peak}$ & $t_{j_{\mathrm{peak},r}}$ & Peak epoch in $r$ band \\
$i\_\text{time\_of\_peak}$ & $t_{j_{\mathrm{peak},i}}$ & Peak epoch in $i$ band \\
$z\_\text{time\_of\_peak}$ & $t_{j_{\mathrm{peak},z}}$ & Peak epoch in $z$ band \\
$\text{time\_span}$ & $\max_j t_j - \min_j t_j$ & Observational duration (not rise/decline time) \\
\bottomrule
\end{tabular}
\end{table*}

\section*{Acknowledgements}
The author thanks Arpita for her continued encouragement and thoughtful feedback during the development of this work. The author also acknowledges the public availability of the SPCC dataset and the open-source scientific Python ecosystem that made this analysis possible.

\section*{Funding}
This research received no specific grant from any funding agency in the public, commercial, or not-for-profit sectors.

\section*{Author Contributions}
Anurag Garg conceived the study, designed the methodology, performed the analysis, interpreted the results, and wrote the manuscript.

\section*{Data Availability}
This study uses the Supernova Photometric Classification Challenge (SPCC) dataset. The processed feature tables and analysis scripts used in this work are available in the public repository \url{https://github.com/mranuraggarg/supernovae_classification}. The manuscript-specific materials associated with the present analysis are archived and cited as \cite{GargRepo2026}.

\section*{Code Availability}
The code used for feature extraction, model training, and feature ablation analysis is available at \url{https://github.com/mranuraggarg/supernovae_classification}. The workflow corresponding to the present manuscript is archived and cited as \cite{GargRepo2026}.

\section*{Conflict of Interest}
The author declares no conflict of interest.

\bibliographystyle{apj}
\bibliography{references}

\end{document}